\documentclass{svproc}

\usepackage[T1]{fontenc}
\usepackage[latin9]{inputenc}
\usepackage{float}
\usepackage{calc}
\usepackage{amsmath}
\usepackage{graphicx}
\usepackage{mathtools}
\usepackage[export]{adjustbox}
\usepackage{listings}
\usepackage{amssymb}
\usepackage{flexisym}
\usepackage{xcolor}
\usepackage{wrapfig}
\usepackage{textcase}
\usepackage{soul}
\usepackage{comment}
\usepackage{multirow}
\usepackage{lmodern}
\usepackage[ruled,vlined]{algorithm2e}
\usepackage{tabularx,booktabs}
\usepackage[numbers,sort&compress]{natbib}
\usepackage{tikz}
\usetikzlibrary{positioning}
\usepackage{nameref}
\usepackage{url}

\graphicspath{{./imgs/}}

\SetKwInOut{KwInput}{Input}
\SetKwInOut{KwOutput}{Output}

\begin{document}
\mainmatter              
\title{Fast Hyperspectral Reconstruction for Neutron Computed Tomography Using Subspace Extraction}
\titlerunning{Fast Hyperspectral Reconstruction}  
\author{Mohammad Samin Nur Chowdhury\inst{1,a} \and Diyu Yang\inst{2,b},
Shimin Tang\inst{3,c} \and Singanallur V. Venkatakrishnan\inst{4,d} \and Andrew W. Needham\inst{5,e} \and Hassina Z. Bilheux\inst{3,f} \and Gregery T. Buzzard\inst{6,g} \and Charles A. Bouman\inst{1,h}}
\authorrunning{M. S. N. Chowdhury et al.} 
\institute{ECE, Purdue University, West Lafayette, IN 47907, USA 
\and
Apple Inc., Cupertino, CA 95014, USA
\and
Neutron Scattering Division, ORNL, Oak Ridge, TN 37830, USA
\and
Electrical \& Engineering Infrastructure Div., ORNL, Oak Ridge, TN 37831, USA
\and
NASA Goddard Space Flight Center, Greenbelt, MD 20771, USA
\and
Department of Mathematics, Purdue University, West Lafayette, IN 47907, USA
\\
$^a$\email{chowdh31@purdue.edu}, $^b$\email{yang1467@purdue.edu}, $^c$\email{tangs@ornl.gov}, $^d$\email{venkatakrisv@ornl.gov}, $^e$\email{andrew.w.needham@nasa.gov}, $^f$\email{bilheuxhn@ornl.gov}, $^g$\email{buzzard@purdue.edu}, $^h$\email{bouman@purdue.edu}
}

\maketitle              

\let\thefootnote\relax\footnote{
This manuscript has been authored by UT-Battelle, LLC, under contract DE-AC05-00OR22725 with the US Department of Energy (DOE). The US government retains and the publisher, by accepting the article for publication, acknowledges that the US government retains a nonexclusive, paid-up, irrevocable, worldwide license to publish or reproduce the published form of this manuscript, or allow others to do so, for US government purposes. DOE will provide public access to these results of federally sponsored research in accordance with the DOE Public Access Plan (http://energy.gov/downloads/doe-public-access-plan).}
\begin{abstract}
Hyperspectral neutron computed tomography enables 3D non-destructive imaging of the spectral characteristics of materials.
In traditional hyperspectral reconstruction, the data for each neutron wavelength bin is reconstructed separately.
This per-bin reconstruction is extremely time-consuming due to the typically large number of wavelength bins.
Furthermore, these reconstructions may suffer from severe artifacts due to the low signal-to-noise ratio in each wavelength bin.

We present a novel fast hyperspectral reconstruction algorithm for computationally efficient and accurate reconstruction of hyperspectral neutron data.
Our algorithm uses a subspace extraction procedure that transforms hyperspectral data into low-dimensional data within an intermediate subspace.
This step effectively reduces data dimensionality and spectral noise.
High-quality reconstructions are then performed within this low-dimensional subspace.
Finally, the algorithm expands the subspace reconstructions into hyperspectral reconstructions.
We apply our algorithm to measured neutron data and demonstrate that it reduces computation and improves reconstruction quality compared to the conventional approach.

\end{abstract}

\section{Introduction}
\label{sec:introduction}

Neutron computed tomography (nCT) enables volumetric reconstruction of a sample from radiographs recorded by exposing the sample to a neutron source at multiple angles.
nCT has been used to identify foreign materials or contamination inside an object \cite{kim1999composite}, to study the effect of environmental factors on material properties, and to monitor the aging of polymers or other materials \cite{chin2007aging}.
Importantly, nCT provides information that is complementary to traditional X-ray CT because neutrons interact directly with the nucleus rather than the electron cloud \cite{vlassenbroeck2007nctvsxct}.

Hyperspectral neutron computed tomography (HSnCT) is a more advanced technique in which a pulsed neutron source illuminates a sample, and a time-of-flight detector measures the projection images across a range of wavelengths - potentially of the order of a few thousand.
Using HSnCT, it is possible to analyze material characteristics like crystallographic phases \cite{woracek20143d} and isotopic compositions \cite{balke2021epithermal}.

\begin{figure}[b!]
\centering
\centerline{\includegraphics[width=0.85\linewidth]{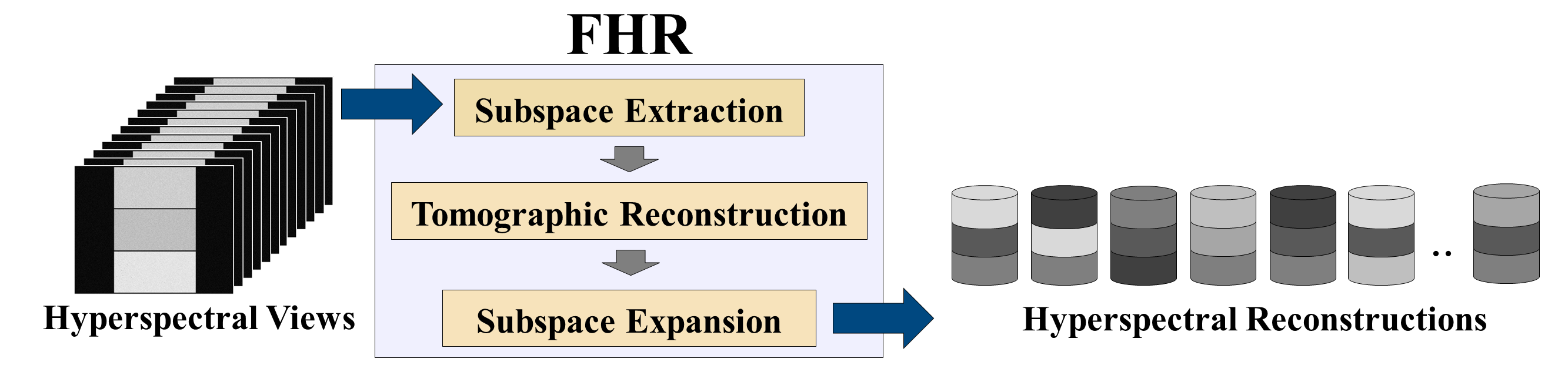}}
\caption{Overview of the fast hyperspectral reconstruction (FHR) algorithm.
FHR reduces the spectral dimension during subspace extraction and restores it during subspace expansion. In between, FHR performs tomographic reconstruction to transition from the sinogram domain to the spatial domain.}
\label{fig:overview}
\end{figure}

HSnCT reconstruction is challenging since data from potentially thousands of neutron wavelength bins must be reconstructed into individual 3D volumes.
Direct hyperspectral reconstruction (DHR) \cite{daugherty2023hyperrecon} is the most established solution to this problem.
In DHR, data from each wavelength bin is reconstructed separately, typically using filtered back projection (FBP) due to its computational speed.
However, even with FBP, DHR is still highly time-consuming since it may require thousands of 3D tomographic reconstructions.
Moreover, producing high-quality reconstructions with FBP requires a large number of projections (also referred to as views) with high signal-to-noise ratios (SNR), which are extremely difficult to acquire due to time and resource constraints.
Consequently, DHR reconstructions using FBP often suffer from severe noise and reconstruction artifacts.
While DHR can alternatively use an advanced algorithm like model-based iterative reconstruction (MBIR) \cite{bouman2022mbir} to improve quality, the computation time would be too high to make it practically implementable.

In this paper, we present a fast hyperspectral reconstruction (FHR) algorithm for effective and efficient reconstruction of HSnCT data.
As shown in Figure~\ref{fig:overview}, the first step of FHR is a subspace extraction procedure \cite{chowdhury2023amd} that transforms the high-dimensional hyperspectral measurements into low-dimensional subspace projection views.
FHR then performs reconstruction within this low-dimensional subspace.
Finally, FHR expands the subspace reconstructions into hyperspectral reconstructions.

The use of subspace extraction in our algorithm serves two purposes:
\begin{itemize}
\item Substantially improves SNR in the reconstructions by approximating the data within a low-dimensional subspace.
\item Significantly accelerates computation by reducing the required number of tomographic reconstructions.
\end{itemize}

We apply our algorithm to measured neutron data collected at the Oak Ridge National Laboratory Spallation Neutron Source SNAP beamline and demonstrate that it is substantially faster and yields better results compared to the baseline DHR method.

\section{Hyperspectral Imaging System}
\label{sec:Imaging_System}

\begin{figure*}[b!]
\centering
\centerline{\includegraphics[width=\linewidth]{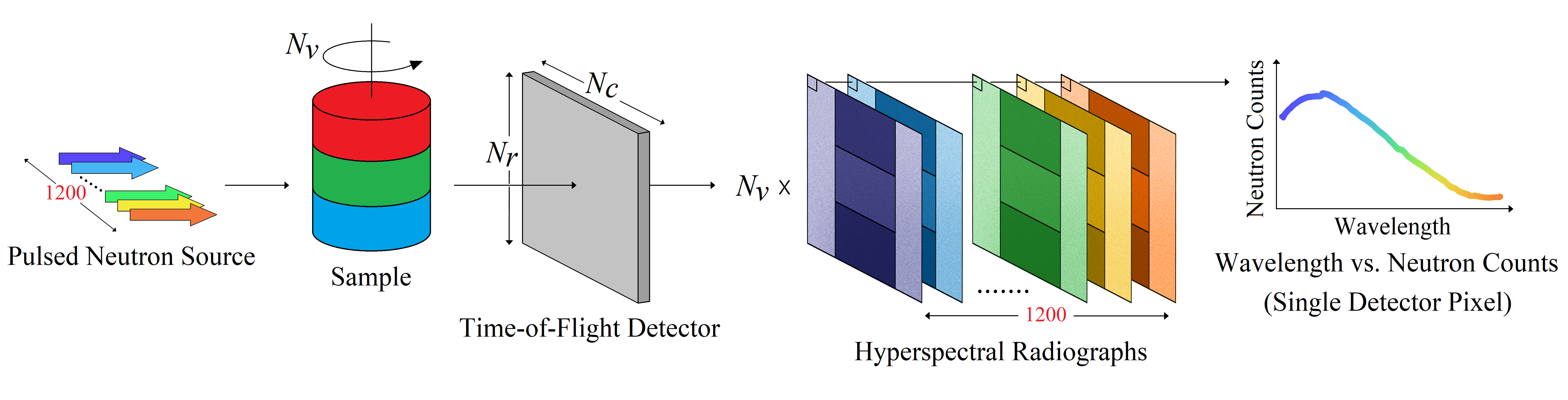}}
\caption{Illustration of a hyperspectral imaging setup at a spallation neutron source.
The arrows on the left represent a pulsed neutron source, which generates a beam of neutrons across a range of wavelengths.
The beam travels through the sample to a $N_r \times N_c$ time-of-flight (TOF) detector, which produces wavelength-resolved radiographs for $N_k=1200$ bins.
The right plot shows neutron counts across the wavelength bins for a single detector pixel.
Such data are collected for $N_v$ orientations of the sample.}
\label{fig:img_sys}
\end{figure*}

Figure~\ref{fig:img_sys} illustrates a standard hyperspectral neutron imaging system that is used to collect wavelength-resolved hyperspectral data from multiple orientations of the sample at a pulsed neutron source \cite{ nelson2018sns}.
The pulse of neutrons passes through the sample and is detected by a 2D time-of-flight (ToF) imaging array \cite{tremsin2012tof}.
The ToF detector counts the number of neutrons at each pixel and for each time interval bin.
These time interval bins then correspond to each neutron's velocity or wavelength.
The specific relationship between the neutron time of flight, $\Delta t$, and the wavelength, $\lambda$, is given by
\begin{equation} \label{eq:wave_tof}
    \lambda = \frac{h}{m_n} \frac{\Delta t}{L} \ ,
\end{equation}
where $h$ is Planck's constant, $m_n$ is the neutron mass, and $L$ is the distance between the source and the detector.

We now introduce the following notation to describe the relevant quantities:
\begin{itemize}
    \item $N_r$ is the number of detector rows
    \item $N_c$ is the number of detector columns
    \item $N_k$ is the number of wavelength bins
    \item $N_v$ is the number of tomographic views
\end{itemize}

The output of the ToF detector is a hyperspectral neutron radiograph in the form of a $N_r \times N_c \times N_k$ array.
In our experiment, we used $N_r=N_c=512$ and $N_k=1200$.
So, a single hyperspectral radiograph in our experiment contains approximately $300$ megapixels.

In order to perform a tomographic scan, the object is rotated to $N_v$ orientations \cite{yang2023orientation, tang2024orientation} and at each orientation, a hyperspectral radiograph, $y_{v,r,c,k}$, is measured where $v,r,c,k$ represent the discrete view, row, column, and wavelength indices.
In addition, a single hyperspectral radiograph is measured with the object removed, which we denote by $y^o_{r,c,k}$.
From this, we can compute the normalized hyperspectral projection views $p$ as
\begin{equation}
\label{eq:neglogattenuation}
p_{v,r,c,k} = -\log \left( \frac {y_{v,r,c,k}} {y_{r,c,k}^o} \right) \ .
\end{equation}
We note that in practice, various corrections must be made in the calculation of \eqref{eq:neglogattenuation} to account for effects like scatter, detector bias, and detector drift \cite{balke2021epithermal}.

\begin{figure*}[b!]
\centering
\centerline{\includegraphics[width=\linewidth]{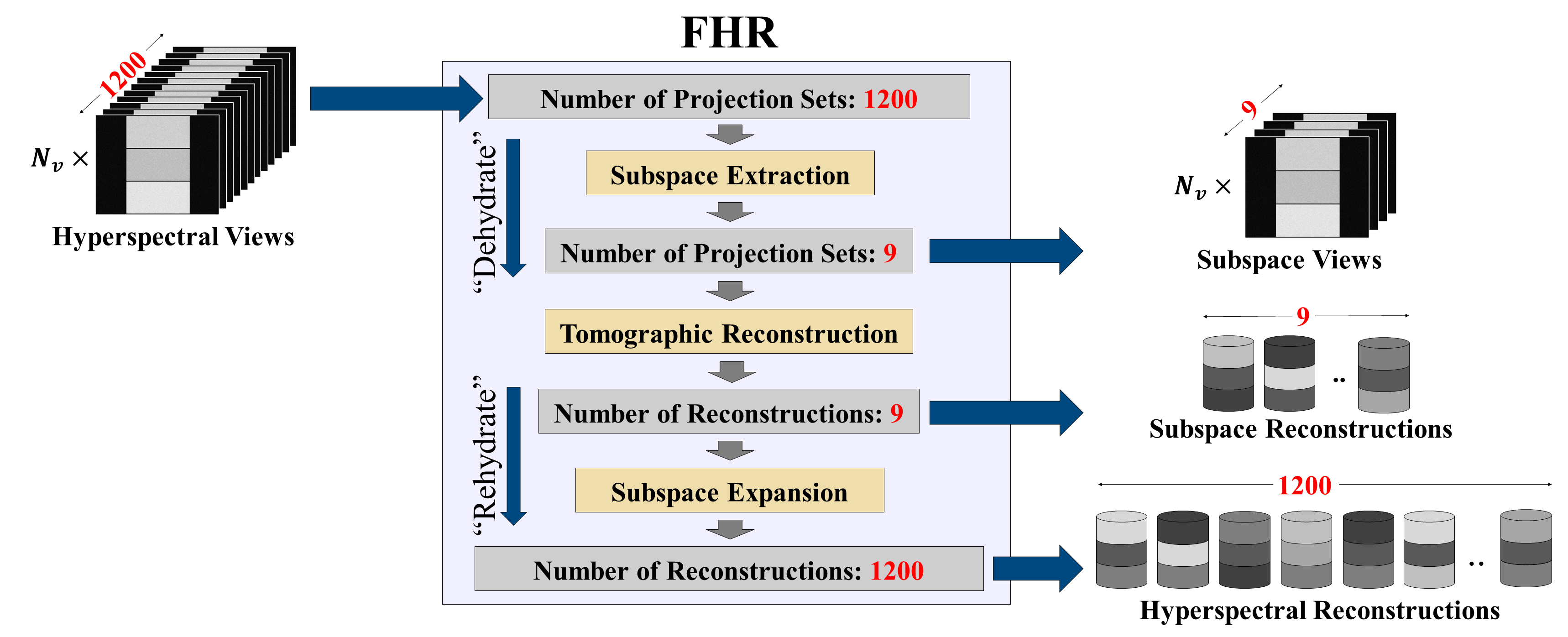}}
\caption{Illustration of the FHR algorithm.
FHR first transforms the $N_k=1200$ dimensional hyperspectral views into $N_s=9$ dimensional subspace views.
Then, it performs tomographic reconstruction to produce 9 subspace reconstructions.
Finally, it expands the 9 subspace reconstructions into 1200 hyperspectral reconstructions.}
\label{fig:FHR_pipeline}
\end{figure*}

\section{Fast Hyperspectral Reconstruction (FHR)}
\label{sec:Fast_Hyper_Recon}

Figure~\ref{fig:FHR_pipeline} illustrates the three steps in FHR consisting of subspace extraction, tomographic reconstruction, and subspace expansion.
FHR reduces the spectral dimension during subspace extraction, then performs volumetric reconstruction in this lower dimensional space, and finally restores the spectral dimension using subspace expansion.
In more colorful terms, the algorithm is ``dehydrating'' the data in the sinogram domain and then ``rehydrating'' in the reconstruction domain.

We use the following notation in the remaining sections:
\begin{itemize}
    \item $N_s$ is the dimension of the subspace
    \item $N_p=N_v \times N_r \times N_c$ is the number of measurements for each wavelength bin
    \item $N_x=N_r \times N_c \times N_c$ is the number of voxels for each wavelength bin
\end{itemize}

\vspace{0.3cm}

\bigskip
\noindent
{\em Subspace Extraction:}

\begin{figure}[b!]
\centering
\includegraphics[width=\linewidth]{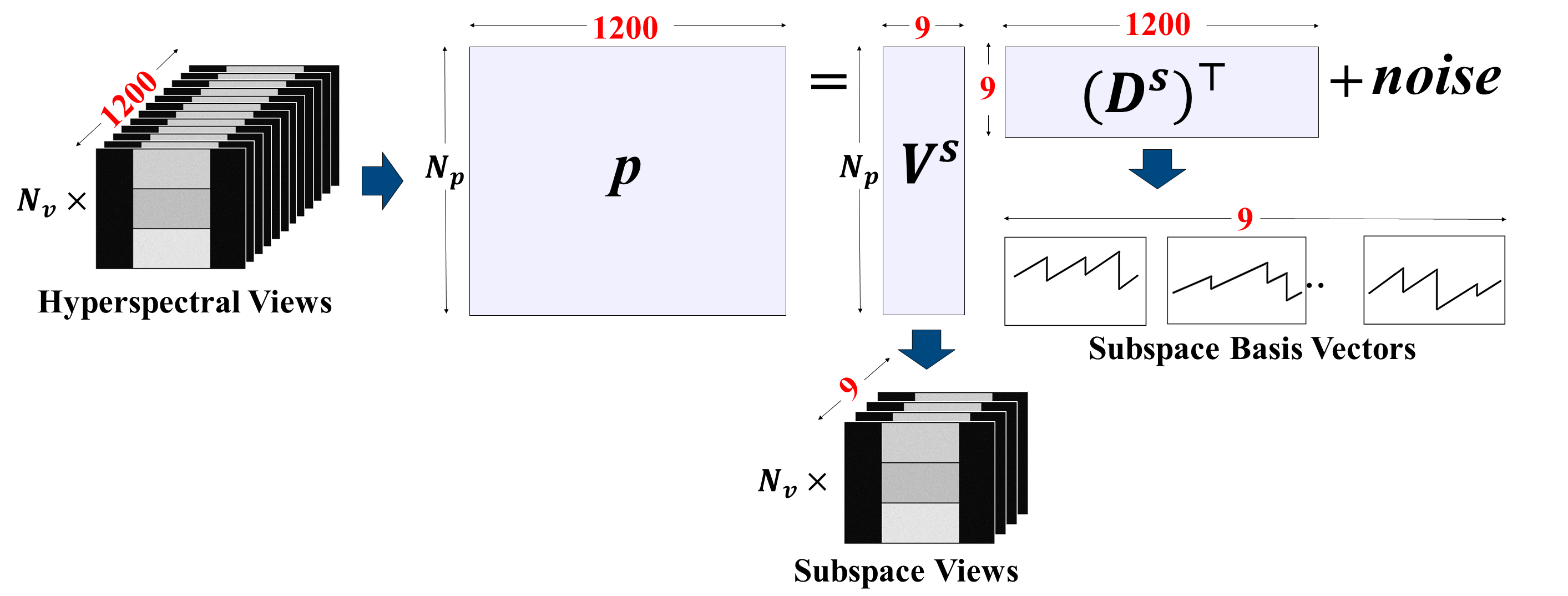}
\caption{Subspace extraction:
FHR employs NMF to decompose the $N_k=1200$ dimensional hyperspectral views ($p$) into $N_s=9$ dimensional subspace views ($V^s$) and corresponding subspace basis vectors ($D^s$).
This approach effectively reduces data dimensions and also eliminates significant spectral noise.}
\label{fig:sub_ext}
\end{figure}

Figure~\ref{fig:sub_ext} illustrates the subspace extraction step of FHR in which the high-dimensional hyperspectral views $p \in \mathbb{R}^{N_p \times N_k}$ are decomposed into low-dimensional subspace views $V^s \in \mathbb{R}^{N_p \times N_s}$ and corresponding subspace basis vectors $D^s \in \mathbb{R}^{N_k \times N_s}$.
The subspace dimension $N_s$ is chosen such that $N_m<N_s<<N_k$, where $N_m$ is the number of materials present in the sample.
The decomposition can be obtained by solving the non-negative matrix factorization (NMF) \cite{pauca2006nmf} problem.

Since $N_s<<N_k$, operations in the subspace domain are much faster than in the hyperspectral domain. 
Also, empirically, the residual difference from the decomposition $\epsilon = p-V^s (D^s)^\top$ is primarily spectral noise.
This greatly reduces the noise in the final reconstructions.

\begin{figure}[h!]
\centering
\includegraphics[width=8cm]{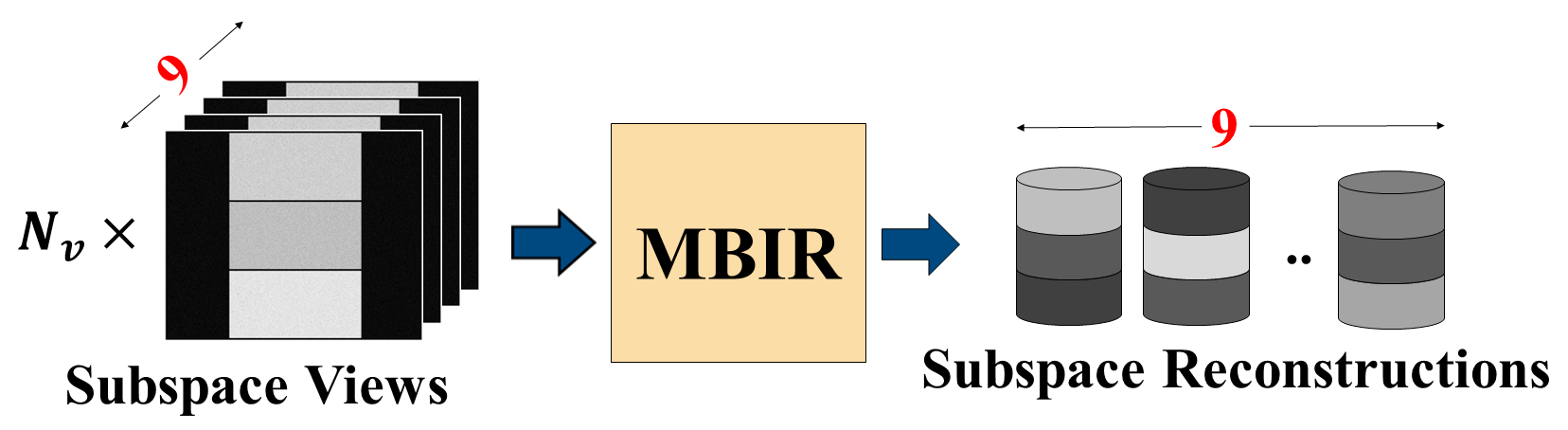}
\caption{Tomographic reconstruction:
FHR computes $N_s=9$ reconstructions from the extracted 9 sets of subspace views using MBIR.}
\label{fig:recon}
\end{figure}

\bigskip
\noindent
{\em Tomographic Reconstruction:}

Figure~\ref{fig:recon} illustrates the reconstruction step of FHR in which $N_s$ reconstructions within the subspace $x^s \in \mathbb{R}^{N_x \times N_s}$ are computed using MBIR \cite{bouman2022mbir}.
MBIR is well known to produce superior reconstruction when dealing with sparse and low SNR measurements - hence is the ideal choice for HSnCT \cite{venkat2021hyperspectral}.
In order to perform MBIR, FHR uses the recently released SVMBIR Python package \cite{svmbir-code}.

We note that MBIR reconstruction tends to be much slower than FBP reconstruction, but since MBIR enables sparse view reconstruction, it can effectively reduce the data acquisition time for HSnCT.
Also, the increased reconstruction time for MBIR is of much less concern since FHR only requires the reconstruction of $N_s$ volumes rather than the $N_k$ 3D volumes required for DHR.

\bigskip
\noindent
{\em Subspace Expansion:}

\begin{figure}[h!]
\centering
\includegraphics[width=\linewidth]{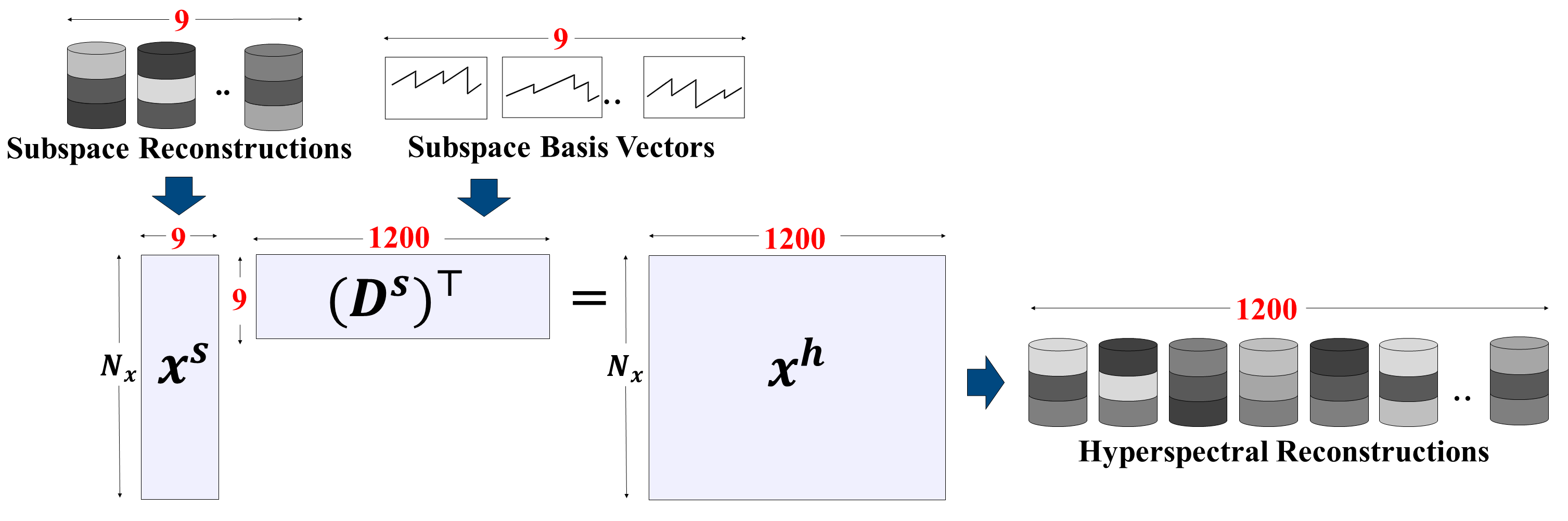}
\caption{Subspace expansion:
FHR expands the $N_s=9$ subspace reconstructions ($x^s$) to $N_k=1200$ hyperspectral reconstructions ($x^h$) using the subspace basis vectors ($D^s$).}
\label{fig:sub_expansion}
\end{figure}

Figure~\ref{fig:sub_expansion} illustrates the subspace expansion step of FHR in which the algorithm expands the subspace reconstructions into hyperspectral reconstructions using $D^s$. 
Since $D^s$ maps each voxel from subspace to hyperspectral coordinates, the resulting hyperspectral reconstruction is of size $x^h \in \mathbb{R}^{N_x \times N_k}$.

\section{Results}
\label{sec:results}

\begin{figure}[b!]
\begin{minipage}[a]{0.5\linewidth}
\centering
\includegraphics[width=0.85\linewidth]{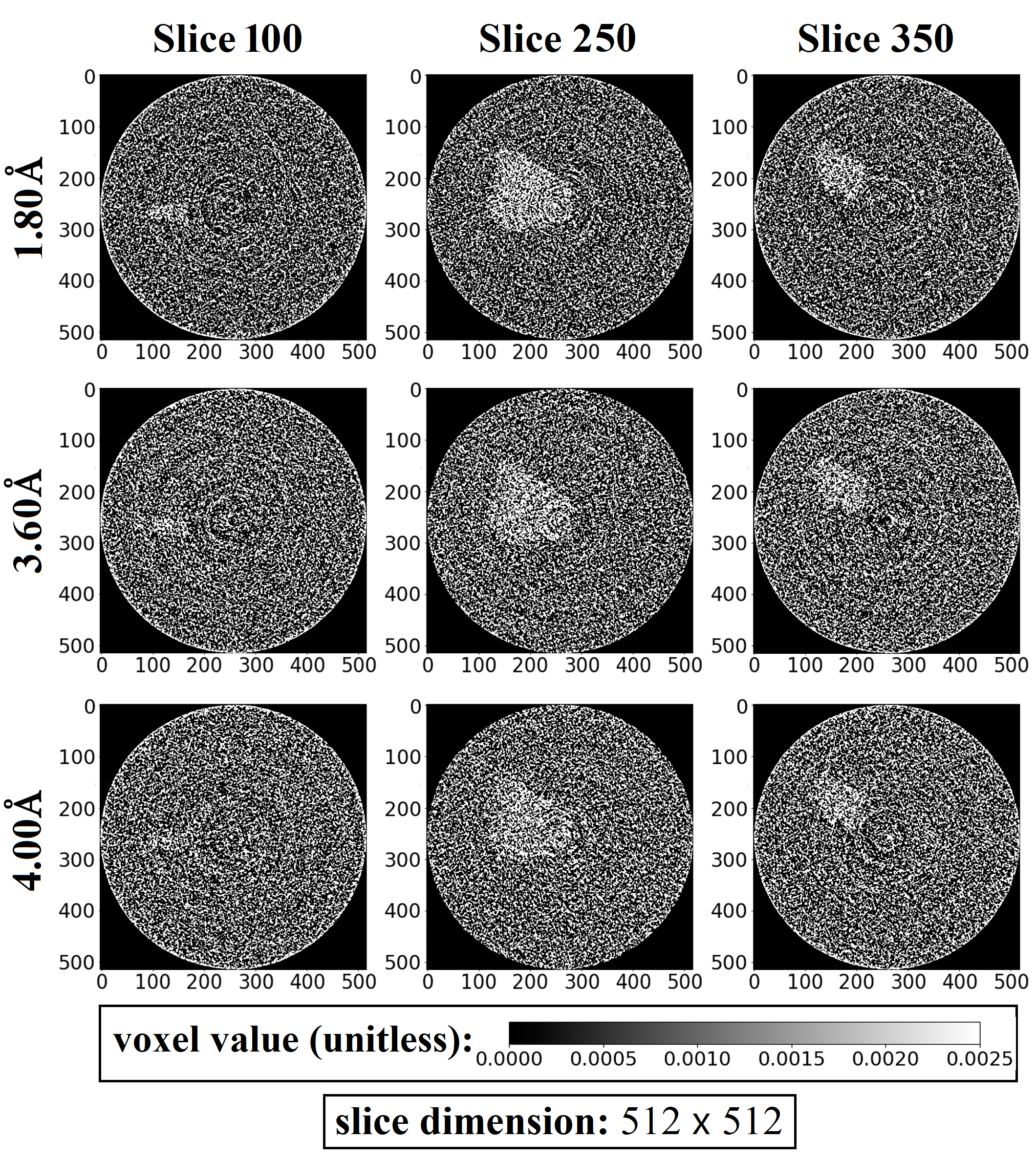}
\centerline{(a) DHR-FBP}\medskip  
\end{minipage}
\hfill
\begin{minipage}[a]{0.5\linewidth}
\centering
\includegraphics[width=0.85\linewidth]{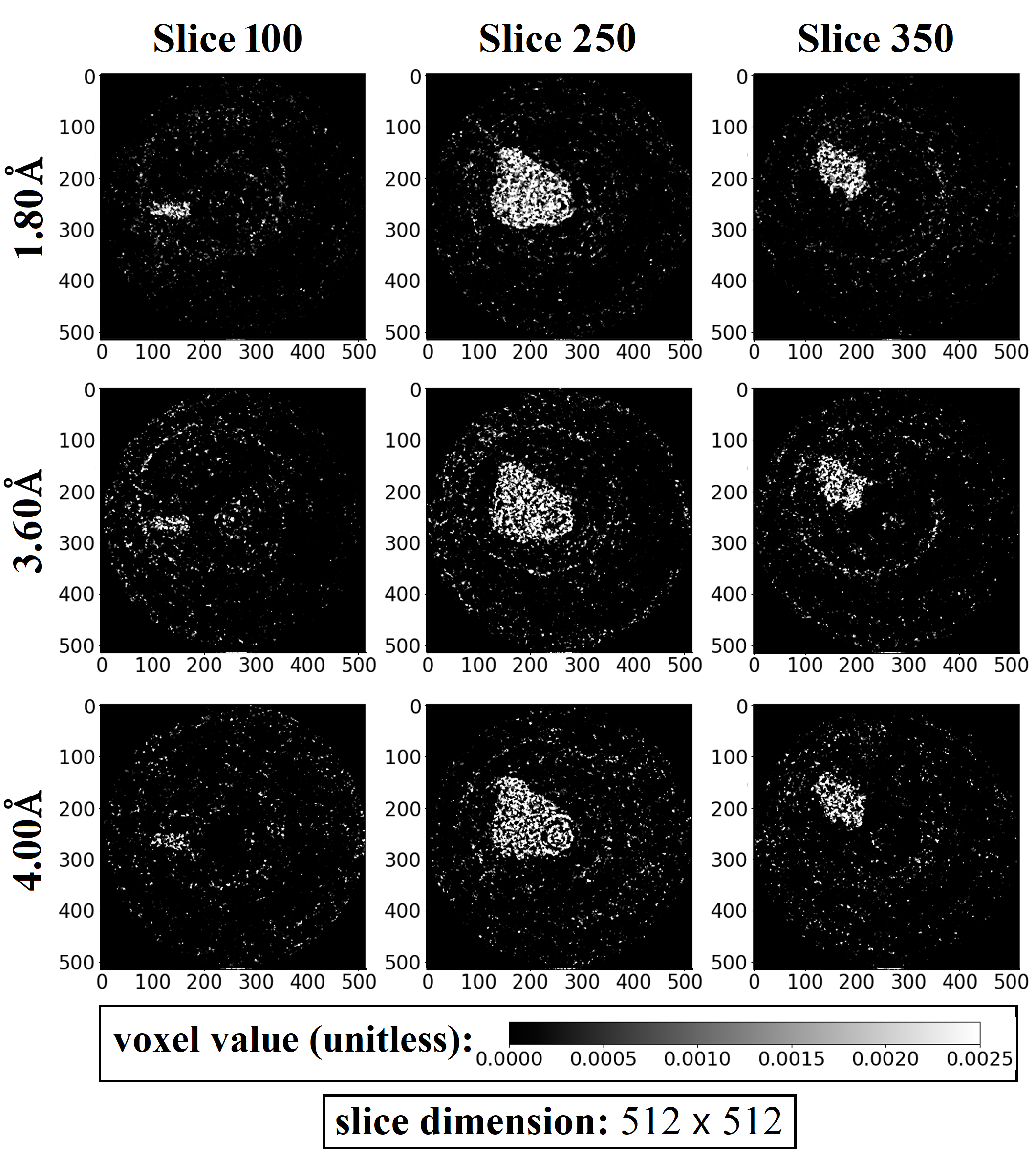}
\centerline{(b) DHR-MBIR}\medskip  
\end{minipage}
\hfill
\begin{minipage}[a]{\linewidth}
\centering
\includegraphics[width=0.425\linewidth]{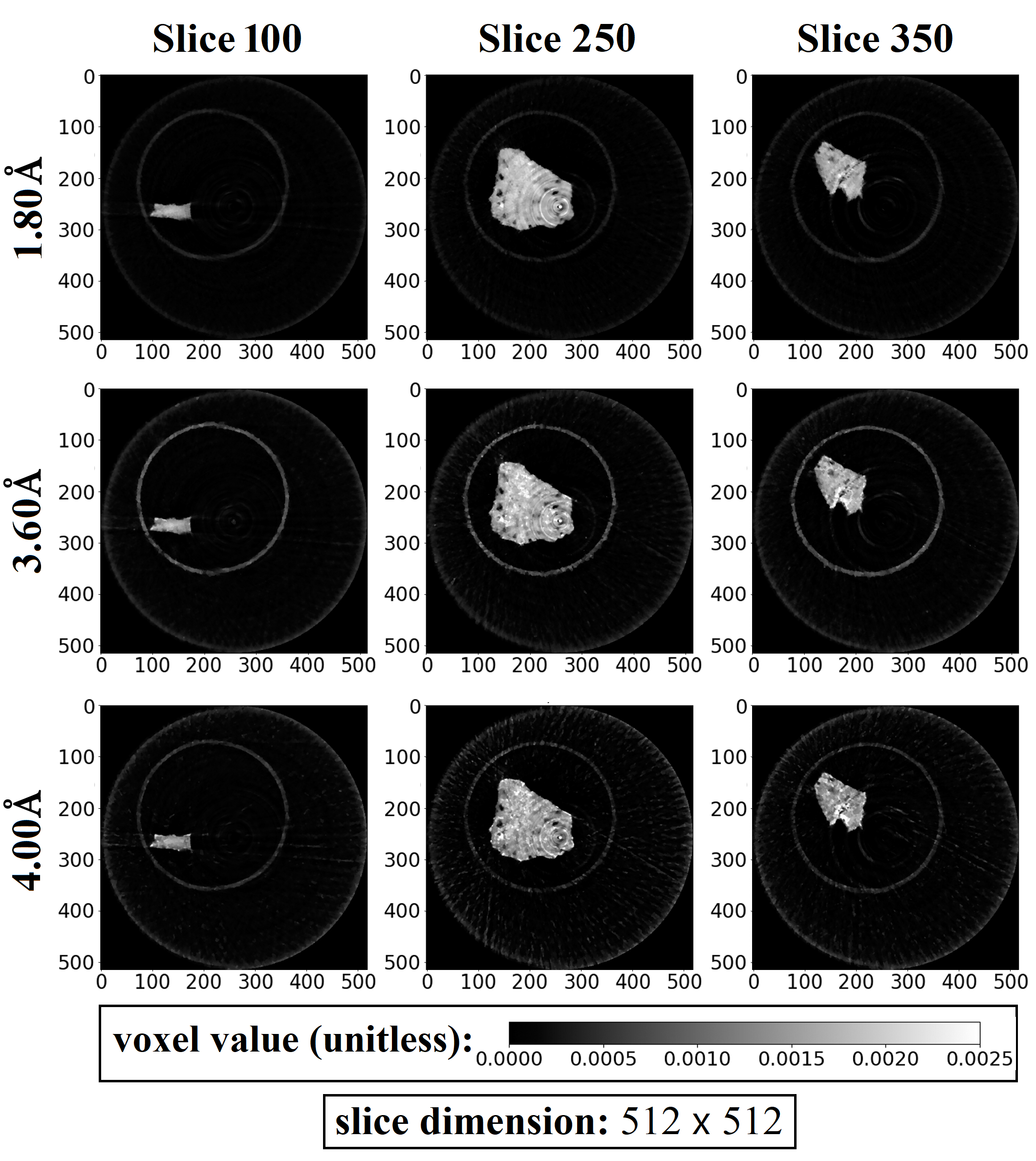}
\centerline{(c) FHR}\medskip
\end{minipage}
\caption{Hyperspectral reconstruction of a moon-rock sample: 
(a) using the baseline DHR method with FBP, (b) using the baseline DHR method with MBIR, and (c) using the proposed FHR algorithm.
The images shown are three different slices from the 3D volumes at three different wavelength bins.
While DHR-MBIR reconstructions exhibit significantly fewer artifacts compared to DHR-FBP, they are still impacted by noise due to the low SNR per wavelength bin. In contrast, the proposed FHR dramatically improves the reconstruction quality by suppressing noise while preserving edge details.}
\label{fig:recon_hyper_real}
\end{figure}

Below, we present results for hyperspectral reconstruction of a Moon rock sample from Apollo Mission 14, measured at the Spallation Neutron Source at Oak Ridge National Laboratory.
HSnCT data was collected with $N_v=53$ views, $N_r=N_c=512$, and $N_k=1200$ hyperspectral wavelength bins.
So, the resulting 3D volumes have dimensions $512\times 512\times 512$.
We heuristically selected $N_s=9$ as the dimension of the subspace.

Figure~\ref{fig:recon_hyper_real} presents hyperspectral reconstructions for the measured data using DHR-FBP, DHR-MBIR, and the proposed FHR algorithm.
Notice that while DHR-MBIR reconstructions demonstrate superior quality compared to DHR-FBP, they still contain significantly more noise and artifacts than FHR reconstructions.
This is because of the initial subspace extraction in FHR, which removed the spectral noise orthogonal to the $N_s=9$ dimensional space.
Table~\ref{table:hr_real} shows a quantitative performance comparison among DHR-FBP, DHR-MBIR, and FHR.
For this dataset, FHR resulted in an SNR approximately 32 dB higher than DHR-FBP and 8.5 dB higher than DHR-MBIR.
Moreover, FHR was over 10 times faster than DHR-FBP and over 100 times faster than DHR-MBIR, highlighting the computational efficiency of our approach.

\begin{table}[h!]
\begin{center}
\caption{Quantitative performance comparison among DHR-FBP, DHR-MBIR, and the proposed FHR algorithm.}
\label{table:hr_real}
\begin{tabular}{|c|c|c|}
  \hline
  Algorithm & SNR & Computation Time\\
  \hline
  DHR-FBP & -6.18 dB & 557.03 min\\
  \hline
  DHR-MBIR & 17.40 dB & 6181.84 min\\
  \hline
  FHR & 25.94 dB & 50.87 min\\
  \hline
\end{tabular}
\end{center}
\end{table}

\section{Conclusion}
\label{sec:conclusion}

We present a fast hyperspectral reconstruction (FHR) algorithm for HSnCT that produces higher-quality reconstructions with dramatically less computation when compared to a conventional direct HSnCT reconstruction method.
The intermediate subspace extraction introduced in FHR decreases data dimensionality by a factor of up to 100 and reduces spectral noise.
In our experiments, this resulted in a decrease in overall computational time by factors of 10 and 100 compared to DHR-FBP and DHR-MBIR, respectively.
We also observed an increase in reconstruction SNR by approximately 32 dB over DHR-FBP and 8.5 dB over DHR-MBIR.

\section*{Acknowledgment}
\label{sec:acknowledgment}

C. Bouman was partially supported by the Showalter Trust. 
This research used resources at the Spallation Neutron Source, a DOE Office of Science User Facility operated by the Oak Ridge National Laboratory.
The beam time was allocated to the Spallation Neutrons and Pressure Diffractometer (SNAP) instrument on proposal number IPTS-25265.

%
%

\end{document}